# A measurement of the mean central optical depth of galaxy clusters via the pairwise kinematic Sunyaev-Zel'dovich effect with SPT-3G and DES

E. Schiappucci,[1] F. Bianchini,[2,3,4] M. Aguena,[5] M. Archipley,[6,7] L. Balkenhol,[1] L. E. Bleem,[8,9] P. Chaubal,[1] T. M. Crawford,[9,10] S. Grandis,[11,12] Y. Omori,[13,14,15,16] C. L. Reichardt,[1] E. Rozo,[17] E. S. Rykoff,[15,18] C. To,[19] T. M. C. Abbott,[20] P. A. R. Ade,[21] O. Alves,[22,23] A. J. Anderson,[24,9,10] F. Andrade-Oliveira,[22] J. Annis,[25] J. S. Avva,[26] D. Bacon,[27] K. Benabed,[28] A. N. Bender,[8,9] B. A. Benson,[24,9,10] G. M. Bernstein,[29] E. Bertin,[30,31] S. Bocquet,[32] F. R. Bouchet,[28] D. Brooks,[33] D. L. Burke,[16,18] J. E. Carlstrom,[9,34,35,8,10] A. Carnero Rosell,[36,23,37] M. Carrasco Kind,[38,6] J. Carretero,[39] T. W. Cecil,[8] C. L. Chang,[8,9,10] P. M. Chichura,[35,9] T.-L. Chou,[35,9] M. Costanzi,[40,41,42] A. Cukierman,[2,4,3] L. N. da Costa,[23] C. Daley,[6] T. de Haan,[43] S. Desai,[44] K. R. Dibert,[10,9] H. T. Diehl,[25] M. A. Dobbs,[45,46] P. Doel,[33] C. Doux,[35,9] D. Dutcher,[35,9] S. Everett,[47] W. Everett,[48] C. Feng,[49] K. R. Ferguson,[50] I. Ferrero,[51] A. Ferté,[47] B. Flaugher,[25] A. Foster,[52] J. Frieman,[25,15] S. Galli,[28] A. E. Gambrel,[9] J. García-Bellido,[53] R. W. Gardner,[34] M. Gatti,[29] T. Giannantonio,[54,55] N. Goeckner-Wald,[3,2] D. Gruen,[32] R. Gualtieri,[8] S. Guns,[26] G. Gutierrez,[25] N. W. Halverson,[48,56] S. R. Hinton,[57] E. Hivon,[28] G. P. Holder,[58] D. L. Hollowood,[59] W. L. Holzapfel,[26] K. Honscheid,[19,60] J. C. Hood,[9] N. Huang,[26] D. J. James,[61] L. Knox,[62] M. Korman,[52] K. Kuehn,[63,64] C.-L. Kuo,[2,3,4] O. Lahav,[33] A. T. Lee,[26,65] C. Lidman,[66,67] M. Lima,[68,23] A. E. Lowitz,[9] C. Lu,[58] M. March,[29] J. Mena-Fernández,[69] F. Menanteau,[38,6] M. Millea,[26] R. Miquel,[70,39] J. J. Mohr,[71,32] J. Montgomery,[45] J. Muir,[72] T. Natoli,[9] G. I. Noble,[45] V. Novosad,[73] R. L. C. Ogando,[74] S. Padin,[9,75] Z. Pan,[8,9,35] F. Paz-Chinchón,[38,54] M. E. S. Pereira,[76] A. Pieres,[23,74] A. A. Plazas Malagón,[77] K. Prabhu,[62] J. Prat,[13,15] W. Quan,[35,9] A. Rahlin,[24,9] M. Raveri,[29] M. Rodriguez-Monroy,[78] A. K. Romer,[78] M. Rouble,[45] J. E. Ruhl,[52] E. Sanchez,[69] V. Scarpine,[25] M. Schubnell,[22] G. Smecher,[79] M. Smith,[80] M. Soares-Santos,[22] J. A. Sobrin,[35,9] E. Suchyta,[81] A. Suzuki,[65] G. Tarle,[22] D. Thomas,[27] K. L. Thompson,[2,3,4] B. Thorne,[62] C. Tucker,[21] C. Umilta,[58] J. D. Vieira,[6,58,7] M. Vincenzi,[27,80] G. Wang,[8] N. Weaverdyck,[22,82] J. Weller,[71,83] N. Whitehorn,[84] W. L. K. Wu,[2,4] V. Yefremenko,[8] and M. R. Young[24,9]

(SPT-3G and DES Collaborations)

[1]*School of Physics, University of Melbourne, Parkville, VIC 3010, Australia*
[2]*Kavli Institute for Particle Astrophysics and Cosmology,
Stanford University, 452 Lomita Mall, Stanford, CA, 94305, USA*
[3]*Department of Physics, Stanford University, 382 Via Pueblo Mall, Stanford, CA, 94305, USA*
[4]*SLAC National Accelerator Laboratory, 2575 Sand Hill Road, Menlo Park, CA, 94025, USA*
[5]*Laboratório Interinstitucional de e-Astronomia-LIneA,
Rua Gal. José Cristino 77, Rio de Janeiro, RJ—20921-400, Brazil*
[6]*Department of Astronomy, University of Illinois Urbana-Champaign,
1002 West Green Street, Urbana, IL, 61801, USA*
[7]*Center for AstroPhysical Surveys, National Center for Supercomputing Applications, Urbana, IL, 61801, USA*
[8]*High-Energy Physics Division, Argonne National Laboratory,
9700 South Cass Avenue., Lemont, IL, 60439, USA*
[9]*Kavli Institute for Cosmological Physics, University of Chicago,
5640 South Ellis Avenue, Chicago, IL, 60637, USA*
[10]*Department of Astronomy and Astrophysics, University of Chicago,
5640 South Ellis Avenue, Chicago, IL, 60637, USA*
[11]*University Observatory, Faculty of Physics, Ludwig-Maximilians-Universität, Scheinerstr. 1, 81679 Munich, Germany*
[12]*Excellence Cluster ORIGINS, Boltzmannstr. 2, 85748, Garching, Germany*
[13]*Department of Astronomy and Astrophysics, University of Chicago, Chicago, IL 60637, USA*
[14]*Department of Physics, Stanford University, 382 Via Pueblo Mall, Stanford, CA 94305, USA*
[15]*Kavli Institute for Cosmological Physics, University of Chicago, Chicago, IL 60637, USA*
[16]*Kavli Institute for Particle Astrophysics & Cosmology,
P. O. Box 2450, Stanford University, Stanford, CA 94305, USA*
[17]*Department of Physics, University of Arizona, Tucson, AZ 85721, USA*
[18]*SLAC National Accelerator Laboratory, Menlo Park, CA 94025, USA*
[19]*Center for Cosmology and Astro-Particle Physics,
The Ohio State University, Columbus, OH 43210, USA*
[20]*Cerro Tololo Inter-American Observatory, NSF's National Optical-Infrared
Astronomy Research Laboratory, Casilla 603, La Serena, Chile*
[21]*School of Physics and Astronomy, Cardiff University, Cardiff CF24 3YB, United Kingdom*
[22]*Department of Physics, University of Michigan, Ann Arbor, MI 48109, USA*
[23]*Laboratório Interinstitucional de e-Astronomia - LIneA,
Rua Gal. José Cristino 77, Rio de Janeiro, RJ - 20921-400, Brazil*

[24] *Fermi National Accelerator Laboratory, MS209, P.O. Box 500, Batavia, IL, 60510, USA*
[25] *Fermi National Accelerator Laboratory, P. O. Box 500, Batavia, IL 60510, USA*
[26] *Department of Physics, University of California, Berkeley, CA, 94720, USA*
[27] *Institute of Cosmology and Gravitation, University of Portsmouth, Portsmouth, PO1 3FX, UK*
[28] *Institut d'Astrophysique de Paris, UMR 7095, CNRS & Sorbonne Université, 98 bis boulevard Arago, 75014 Paris, France*
[29] *Department of Physics and Astronomy, University of Pennsylvania, Philadelphia, PA 19104, USA*
[30] *CNRS, UMR 7095, Institut d'Astrophysique de Paris, F-75014, Paris, France*
[31] *Sorbonne Universités, UPMC Univ Paris 06, UMR 7095, Institut d'Astrophysique de Paris, F-75014, Paris, France*
[32] *University Observatory, Faculty of Physics, Ludwig-Maximilians-Universität, Scheinerstr. 1, 81679 Munich, Germany*
[33] *Department of Physics & Astronomy, University College London, Gower Street, London, WC1E 6BT, UK*
[34] *Enrico Fermi Institute, University of Chicago, 5640 South Ellis Avenue, Chicago, IL, 60637, USA*
[35] *Department of Physics, University of Chicago, 5640 South Ellis Avenue, Chicago, IL, 60637, USA*
[36] *Instituto de Astrofisica de Canarias, E-38205 La Laguna, Tenerife, Spain*
[37] *Universidad de La Laguna, Dpto. Astrofísica, E-38206 La Laguna, Tenerife, Spain*
[38] *Center for Astrophysical Surveys, National Center for Supercomputing Applications, 1205 West Clark St., Urbana, IL 61801, USA*
[39] *Institut de Física d'Altes Energies (IFAE), The Barcelona Institute of Science and Technology, Campus UAB, 08193 Bellaterra (Barcelona) Spain*
[40] *Astronomy Unit, Department of Physics, University of Trieste, via Tiepolo 11, I-34131 Trieste, Italy*
[41] *INAF-Osservatorio Astronomico di Trieste, via G. B. Tiepolo 11, I-34143 Trieste, Italy*
[42] *Institute for Fundamental Physics of the Universe, Via Beirut 2, 34014 Trieste, Italy*
[43] *High Energy Accelerator Research Organization (KEK), Tsukuba, Ibaraki 305-0801, Japan*
[44] *Department of Physics, IIT Hyderabad, Kandi, Telangana 502285, India*
[45] *Department of Physics and McGill Space Institute, McGill University, 3600 Rue University, Montreal, Quebec H3A 2T8, Canada*
[46] *Canadian Institute for Advanced Research, CIFAR Program in Gravity and the Extreme Universe, Toronto, ON, M5G 1Z8, Canada*
[47] *Jet Propulsion Laboratory, California Institute of Technology, 4800 Oak Grove Dr., Pasadena, CA 91109, USA*
[48] *CASA, Department of Astrophysical and Planetary Sciences, University of Colorado, Boulder, CO, 80309, USA*
[49] *Department of Astronomy, School of Physical Sciences, University of Science and Technology of China, Hefei, Anhui 230026, China*
[50] *Department of Physics and Astronomy, University of California, Los Angeles, CA, 90095, USA*
[51] *Institute of Theoretical Astrophysics, University of Oslo. P.O. Box 1029 Blindern, NO-0315 Oslo, Norway*
[52] *Department of Physics, Case Western Reserve University, Cleveland, OH, 44106, USA*
[53] *Instituto de Fisica Teorica UAM/CSIC, Universidad Autonoma de Madrid, 28049 Madrid, Spain*
[54] *Institute of Astronomy, University of Cambridge, Madingley Road, Cambridge CB3 0HA, UK*
[55] *Kavli Institute for Cosmology, University of Cambridge, Madingley Road, Cambridge CB3 0HA, UK*
[56] *Department of Physics, University of Colorado, Boulder, CO, 80309, USA*
[57] *School of Mathematics and Physics, University of Queensland, Brisbane, QLD 4072, Australia*
[58] *Department of Physics, University of Illinois Urbana-Champaign, 1110 West Green Street, Urbana, IL, 61801, USA*
[59] *Santa Cruz Institute for Particle Physics, Santa Cruz, CA 95064, USA*
[60] *Department of Physics, The Ohio State University, Columbus, OH 43210, USA*
[61] *Center for Astrophysics | Harvard & Smithsonian, 60 Garden Street, Cambridge, MA 02138, USA*
[62] *Department of Physics & Astronomy, University of California, One Shields Avenue, Davis, CA 95616, USA*
[63] *Australian Astronomical Optics, Macquarie University, North Ryde, NSW 2113, Australia*
[64] *Lowell Observatory, 1400 Mars Hill Rd, Flagstaff, AZ 86001, USA*
[65] *Physics Division, Lawrence Berkeley National Laboratory, Berkeley, CA, 94720, USA*
[66] *Centre for Gravitational Astrophysics, College of Science, The Australian National University, ACT 2601, Australia*
[67] *The Research School of Astronomy and Astrophysics, Australian National University, ACT 2601, Australia*
[68] *Departamento de Física Matemática, Instituto de Física, Universidade de São Paulo, CP 66318, São Paulo, SP, 05314-970, Brazil*
[69] *Centro de Investigaciones Energéticas, Medioambientales y Tecnológicas (CIEMAT), Madrid, Spain*
[70] *Institució Catalana de Recerca i Estudis Avançats, E-08010 Barcelona, Spain*
[71] *Max Planck Institute for Extraterrestrial Physics, Giessenbachstrasse, 85748 Garching, Germany*
[72] *Perimeter Institute for Theoretical Physics, 31 Caroline St. North, Waterloo, ON N2L 2Y5, Canada*
[73] *Materials Sciences Division, Argonne National Laboratory, 9700 South Cass Avenue, Lemont, IL, 60439, USA*

[74] Observatório Nacional, Rua Gal. José Cristino 77, Rio de Janeiro, RJ - 20921-400, Brazil
[75] California Institute of Technology, 1200 East California Boulevard., Pasadena, CA, 91125, USA
[76] Hamburger Sternwarte, Universität Hamburg, Gojenbergsweg 112, 21029 Hamburg, Germany
[77] Department of Astrophysical Sciences, Princeton University, Peyton Hall, Princeton, NJ 08544, USA
[78] Department of Physics and Astronomy, Pevensey Building, University of Sussex, Brighton, BN1 9QH, UK
[79] Three-Speed Logic, Inc., Victoria, B.C., V8S 3Z5, Canada
[80] School of Physics and Astronomy, University of Southampton, Southampton, SO17 1BJ, UK
[81] Computer Science and Mathematics Division, Oak Ridge National Laboratory, Oak Ridge, TN 37831
[82] Lawrence Berkeley National Laboratory, 1 Cyclotron Road, Berkeley, CA 94720, USA
[83] Universitäts-Sternwarte, Fakultät für Physik, Ludwig-Maximilians Universität München, Scheinerstr. 1, 81679 München, Germany
[84] Department of Physics and Astronomy, Michigan State University, East Lansing, MI 48824, USA

We infer the mean optical depth of a sample of optically-selected galaxy clusters from the Dark Energy Survey (DES) via the pairwise kinematic Sunyaev-Zel'dovich (kSZ) effect. The pairwise kSZ signal between pairs of clusters drawn from the DES Year-3 cluster catalog is detected at $4.1\,\sigma$ in cosmic microwave background (CMB) temperature maps from two years of observations with the SPT-3G camera on the South Pole Telescope. After cuts, there are 24,580 clusters in the $\sim 1{,}400$ deg$^2$ of the southern sky observed by both experiments. We infer the mean optical depth of the cluster sample with two techniques. The optical depth inferred from the pairwise kSZ signal is $\bar{\tau}_e = (2.97 \pm 0.73) \times 10^{-3}$, while that inferred from the thermal SZ signal is $\bar{\tau}_e = (2.51 \pm 0.55^{\text{stat}} \pm 0.15^{\text{syst}}) \times 10^{-3}$. The two measures agree at $0.6\,\sigma$. We perform a suite of systematic checks to test the robustness of the analysis.

## I. INTRODUCTION

The Sunyaev-Zel'dovich (SZ) effect [1, 2] occurs when free electrons in the hot intracluster medium of galaxy clusters inverse Compton scatter photons of the cosmic microwave background (CMB). The SZ effect is one of the largest sources of secondary CMB anisotropy and enables powerful probes of astrophysics and cosmology [e.g., 3, 4]. The SZ effect is normally subdivided into the thermal and kinematic Sunyaev-Zel'dovich effects. The thermal SZ (tSZ) effect is due to an energy transfer from the hot electrons to the CMB photons, distorting the CMB black body spectrum by shifting photons to higher frequencies. The kinematic SZ (kSZ) effect is due to the bulk velocity of the electrons slightly changing the apparent temperature of the black body spectrum. While the tSZ effect has been measured through its contribution to the CMB power spectrum and bispectrum, and detected at the individual cluster level [5–10], measuring the kSZ effect is more challenging because of its lower amplitude and spectral degeneracy with the CMB temperature fluctuations [11]. However, measuring the kSZ effect is of great interest since it could potentially be used to constrain both cosmological and astrophysical parameters [12–15], particularly breaking the $f - \sigma_8$ degeneracy that other cosmological probes are incapable of resolving [16].

Although the amplitude of the kSZ signal is small, recent studies have detected the effect. The first detection of the kSZ signal was made using high-resolution CMB data from the Atacama Cosmology Telescope (ACT) [17] in conjunction with the Baryon Oscillation Spectroscopic Survey (BOSS) data release 9 spectroscopic galaxy catalog [18]. A pairwise statistical approach was applied to measure the kSZ signal, which takes into account that on average clusters are falling towards each other due to gravity and this gives rise to a signal that can be measured. This technique led to a rejection of the null-signal hypothesis with a p-value of 0.002 [19]. This pairwise approach was also adopted in a similar analysis using data from the SPT-SZ camera on the South Pole Telescope [20, 21] and the Dark Energy Survey (DES) [22] Year 1 cluster catalog [23], which resulted in a $4.2\sigma$ detection of the pairwise signal. This was the first study to probe the kSZ signal using a photometric redshift cluster sample. More recent analyses using newer ACT CMB data combined with the Sloan Digital Sky Survey (SDSS) galaxy catalog [24], and *Planck* collaboration CMB data in conjunction with the Dark Energy Spectroscopic Instrument (DESI) Legacy Imaging survey galaxy catalog [25], reported $>5\sigma$ evidence for the pairwise kSZ signal. These analyses used spectroscopic galaxy catalogs which provide more accurate redshift measurements in comparison with photometric catalogs, a key feature for detecting the pairwise kSZ signal at high significance. Other methods such as the projected fields [26] technique have obtained a $3.8 - 4.2\sigma$ detection of the kSZ effect, meanwhile a velocity reconstruction approach was used to measure the kSZ signal with a $6.5\sigma$ detection [27].

In this work, we use the CMB temperature maps from SPT-3G, the third-generation camera on the SPT, and a cluster catalog from Year 3 DES data (DES-Y3) to probe the pairwise kSZ effect. We achieve this by applying a matched filter to extract the clusters' SZ imprints on temperatures from the CMB maps and then applying a pairwise statistical approach to the catalog. We also test the robustness of the measurement by using different covariance estimation techniques, null tests, and an analysis of systematics. As a final test, we derive the mean optical depth from the tSZ by using a $y - \tau$ scaling relation calibrated on N-body simulations [28], similarly to previous work [29, 30]. Measuring the tSZ simultaneously with kSZ can break the degeneracy with astrophysics of the kSZ effect and thus be very useful to constrain cosmology.

The paper is organized as follows. In Section II, we briefly describe the theory behind the kSZ effect, its connection to the halo pairwise velocity, and the theoretical template used for modeling the expected signal. The DES and SPT datasets used in this analysis are introduced in Section III. In Section IV, we detail the analysis methods that we use to recover the signal from the data, before describing in Section V the set of simulations that we use to verify our pipeline and to estimate the detection significance expected for our datasets. In Section VI, we present our results, compare with simulations, discuss some robustness tests as well as systematics that could affect the observed signal, and the estimation of the mean optical depth of the cluster catalog through the tSZ. Finally, we briefly summarize our results in Section VII, and discuss the main implications for future analyses.

We use the $\Lambda$CDM model with the best-fit *Planck* 2018 [31] TT+TE+EE+lowE+lensing+BAO cosmological parameters to compute theoretical predictions and to translate redshifts into distances: $H_0 = 67.66 \, \text{km s}^{-1} \, \text{Mpc}^{-1}$, $\Omega_c h^2 = 0.11933$, $\Omega_b h^2 = 0.02242$, $\sigma_8 = 0.8102$, $n_s = 0.9665$. [1]

## II. THEORETICAL BACKGROUND

### A. The pairwise kinematic Sunyaev-Zel'dovich effect

The kSZ effect from a galaxy cluster $i$ produces a fractional shift in the CMB temperature $\Delta T/T_{\text{CMB}}$ proportional to the cluster's velocity $\mathbf{v_i}$ along the line of sight $\hat{\mathbf{r}}_i$:

$$\frac{\Delta T}{T_{\text{CMB}}}(\hat{\mathbf{r}}_i) = -\tau_{e,i} \frac{\hat{\mathbf{r}}_i \cdot \mathbf{v}_i}{c}, \quad (1)$$

where $c$ is the speed of light and $\tau_{e,i}$ is the Thomson optical depth for CMB photons traversing the cluster [2]. This expression assumes a single scattering per photon, which is a good assumption at the low optical depth ($\tau_{e,i} \lesssim 0.01$) of most galaxy clusters. A unique property that the kSZ effect has over the generally brighter tSZ effect is that the kSZ effect depends on the bulk momentum of the ionized cluster gas along the line of sight, and thus can enable tests of the cosmological velocity field [32].

---

[1] The cosmological parameters listed are the Hubble parameter, cold dark matter density, baryon density, current root mean square (rms) of the linear matter fluctuations on scales of $8h^{-1}$Mpc, and the spectral index of the primordial scalar fluctuations respectively.

On scales smaller than the homogeneity scale, we expect pairs of galaxy clusters to fall towards one another on average due to their mutual gravitational pull. Through the kSZ effect, two clusters falling towards one another will leave a potentially detectable dipole pattern on the CMB temperature anisotropy [e.g., 33]. This pattern is called the pairwise kSZ (pkSZ) signal. The average pkSZ amplitude $T_{\text{pkSZ}}(r)$ for all the pairs of galaxy clusters at comoving separation $r$ can be related to the mean pairwise velocity $v_{12}(r)$ of the clusters:

$$T_{\text{pkSZ}}(r) \equiv \bar{\tau}_e \frac{v_{12}(r)}{c} T_{\text{CMB}}, \quad (2)$$

where $\bar{\tau}_e$ is the average optical depth of the sample. This equation is valid with two assumptions: i) the internal motion of the cluster does not introduce any sort of bias, and ii) there is no strong correlation between the optical depth and the velocity of the individual clusters [34]. We adopt a sign convention so that clusters falling towards one another will have a negative relative velocity $v_{12}(r)$ and negative $T_{\text{pkSZ}}$ signal.

We can predict the relative velocity as a function of distance, $v_{12}(r)$, for a specific cosmology and theory of gravity from the statistical distribution of the dark matter haloes. The mean pairwise velocity of haloes $v_{12}(r)$ separated by their comoving distance $r = |\vec{r}_2 - \vec{r}_1|$ can be analytically modeled in linear theory in terms of the two-point matter correlation function $\xi(r)$ as [e.g., 35–37]

$$v_{12}(r,a) \approx -\frac{2}{3} a H(a) f(a) r \frac{b \bar{\xi}(r)}{1 + b^2 \xi(r)}, \quad (3)$$

where $a$ is the scale factor, $H(a)$ the Hubble parameter, $f(a) \equiv d \ln D / d \ln a$ is the growth rate (with $D$ being the linear growth factor), $b$ the mass-averaged halo bias, and $\bar{\xi}$ indicates the average of $\xi(r)$ over a comoving sphere of radius $r$.

Equations (2) and (3) highlight that measurements of the pkSZ are sensitive to a combination of both cluster astrophysics, through the optical depth $\bar{\tau}_e$ and halo bias $b$, and cosmology through the Hubble parameter $H(a)$, the growth rate $f$, and the two-point matter correlation function $\xi(r)$. In particular, the dependence on the growth rate $f$ and the matter correlation function $\xi(r)$ makes measurements of the pkSZ signal sensitive to $f\sigma_8^2$. This provides complementary information to other probes such as redshift space distortions, which are primarily sensitive to $f\sigma_8$ [e.g., 16], and hence could be used to probe dark energy and modifications of gravity [e.g, 37–39].

## III. DATA

### A. SPT-3G temperature maps

In this analysis, we use CMB temperature maps from SPT-3G, the third and latest camera installed on the

South Pole Telescope [SPT, 20, 21]. The SPT is a 10-meter telescope located at the Amundsen-Scott South Pole Station in Antarctica. The SPT-3G focal plane consists of $\sim 16,000$ multichroic, polarization-sensitive transition-edge sensor bolometers which operate in three bands at 95 GHz, 150 GHz, and 220 GHz, with an angular resolution of $\sim 1'$ [40]. The main SPT-3G survey field of approximatly 1,500 deg$^2$ extends from -42° to -70° in declination and -50° to 50° in right ascension. This work uses temperature maps from observations made during the winter season (March-September) of 2019 and 2020.

We refer to [41] for a full description of how the time-ordered data (TOD) are converted into maps, but we will provide a succinct description of the procedures below. The TOD for each of the SPT-3G bolometers are filtered to remove low-frequency noise in the scan direction, with a high-pass filter set at $k_x > 500$. The filtered TOD are binned into map pixels with weights based on the TOD noise level, and calibrated such that the map is in CMB fluctuation temperature units. A flat-sky approximation, the Sanson-Flamsteed projection [42, 43], is used for the map with $0.25'$ square pixels. The map noise levels measured in the $3000 < \ell < 5000$ range are 5.0, 3.9, and 14.0 $\mu$K-arcmin for the coadded 95 GHz, 150 GHz, and 220 GHz temperature maps, respectively. At each frequency, the instrument beam is well represented with a Gaussian with full width at half maximum (FWHM) equal to $1.6'$, $1.2'$, and $1.0'$ at 95, 150, and 220 GHz, respectively. Section IV B and Section VI D describe how these multi-frequency temperature maps are used to extract the CMB+kSZ or Compton-$y$ maps.

### B. DES Year-3 redMaPPer cluster catalog

The second data product used in this analysis is a sample of optically-selected galaxy clusters from the first three years of the Dark Energy Survey (DES). DES is a photometric survey that has mapped out $\sim 5000$ deg$^2$ of the southern sky in the optical to near-infrared bands using the Dark Energy Camera [22], mounted on the 4-meter Blanco telescope at Cerro Tololo Observatory in northern Chile. The cluster catalog has been extracted from DES Year-3 observations with the redMaPPer algorithm [44].

The redMaPPer algorithm is a red-sequence based optical cluster-finding algorithm that is calibrated on a cluster sub-sample for which spectroscopic data are available. The outputs from redMaPPer relevant to the present analysis are: i) the cluster's sky position, given by the angular coordinates of the algorithm's best guess for the central galaxy position; ii) the cluster's photometrically estimated redshift; iii) the optical richness estimate $\lambda$, a weighted sum of the membership probabilities, which is a low-scatter proxy for the cluster mass [e.g., 45, 46]. The underlying idea is that galaxy clusters are concentrations of galaxies containing old red stars thought to be caused by the quenching of star formation due to the cluster's environment. Therefore, the algorithm detects candidates by identifying over-densities of luminous red galaxies and iteratively assigns membership and probabilities for each galaxy identified to be part of a cluster candidate to be in the center of the cluster.

The redMaPPer algorithm has been used to produce both a flux-limited and volume-limited sample using DES Y3 data. The volume-limited sample is independent of survey depth and complete above a certain luminosity, while the flux-limited sample contains more high-$z$ clusters detected in the deeper fields in the survey. In this work, we use the flux-limited catalog. This catalogue contains 41,219 (8,712) clusters in the richness range[2] $\tilde{\lambda} > 10\,(20)$ and spans the photo-$z$ range $0.1 < z < 0.95$.

We restrict the baseline cluster sample used in this work to $z \leq 0.8$ to mitigate the degradation at high redshifts of the completeness and photo-$z$ accuracy. We also impose a cutoff at $\tilde{\lambda} \leq 60$ to eliminate the most massive clusters due to concerns about the possibility that the filtering described in Section IV B will not completely remove the contaminating signals from the cluster itself. We also consider cluster samples with alternative richness ranges in Section VI to test the robustness of the analysis and potential systematic biases.

The DES and SPT-3G surveys overlap over a sky area of $\sim 1,400$ deg$^2$, which we show in Fig. 1. We remove any clusters that are less than 1° from the survey edges, which is a conservative choice to enforce an homogeneous depth coverage; or 10' distance from any point sources detected in the SPT-3G map ($\geq 6$ mJy at 150 GHz) to avoid possible contamination from the point sources onto the clusters. These cuts leave 24,580 (5,797) clusters in the richness range $10\,(20) < \tilde{\lambda} < 60$, which translates to a surface density of 17.6 clusters/deg$^2$ for our $10 \leq \tilde{\lambda} \leq 60$ baseline sample, with a mean redshift of $\bar{z} = 0.54\,(0.52)$ and a typical error in the photo-$z$ of $\sigma_z \sim 0.01(1 + z)$ [47]. The redshift distribution and redshift uncertainties of the full cluster sample are shown in Fig. 2.

## IV. ANALYSIS METHODOLOGY

### A. Pairwise kSZ estimator

As in previous measurements [e.g., 19, 23, 24, 29], we implement the pkSZ estimator $\hat{T}_{\mathrm{pkSZ}}(r)$ introduced by [48]. This estimator for the mean pkSZ signal is,

$$\hat{T}_{\mathrm{pkSZ}}(r) = -\frac{\sum_{i<j,r}[T(\hat{\mathbf{n}}_i) - T(\hat{\mathbf{n}}_j)]\,c_{ij}}{\sum_{i<j,r} c_{ij}^2}, \quad (4)$$

---

[2] We apply cuts to the catalogue using the raw galaxy counts $\tilde{\lambda}$, that are related to the optical richness as $\lambda = s\tilde{\lambda}$ where $s$ is a correction factor based on local survey depth, masking, etc. This choice has been shown to yield a cluster sample with more uniform noise properties, see [23] and references therein.

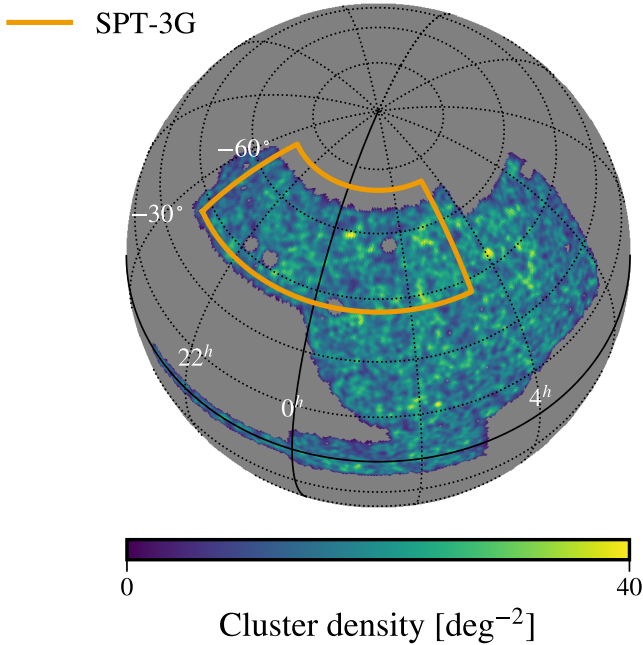

FIG. 1. Number density of the DES Y3 clusters with $10 < \tilde{\lambda} < 60$, smoothed with a Gaussian kernel with $\theta_{\rm FWHM} = 1$ deg for visualization purposes. The solid orange line shows the boundaries of the SPT-3G main survey footprint. These two datasets overlap over approximately 1,400 deg$^2$.

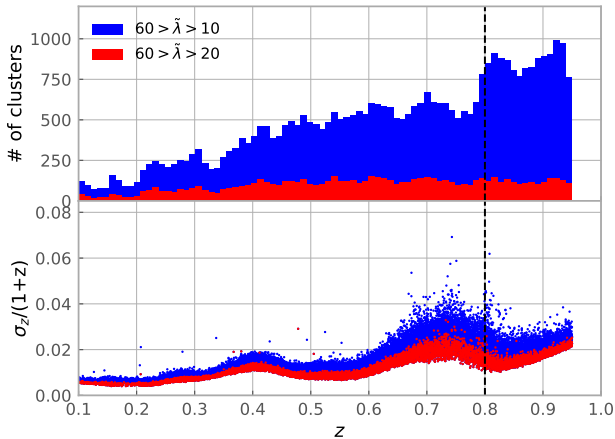

FIG. 2. **Top**: Redshift distribution of the DES Y3 redMaP-Per catalogue for the two richness-based samples. **Bottom**: Photometric redshift errors distribution for the DES Y3 redMaPPer clusters. In each panel, the red color denotes the $20 < \tilde{\lambda} < 60$ sample whereas the blue color refers to our baseline $10 < \tilde{\lambda} < 60$ sample. The vertical dashed black line represents the maximum redshift of clusters that we include in our analysis.

which scales the CMB temperature difference at the two cluster locations (which has an expectation value that depends on the relative velocity between the two clusters due the pkSZ signal) by a geometrical factor, $c_{ij} = \hat{\mathbf{r}}_{ij} \cdot (\hat{\mathbf{r}}_i + \hat{\mathbf{r}}_j)/2$, to account for the projection of the pair separation $\hat{\mathbf{r}}_{ij} = \hat{\mathbf{r}}_i - \hat{\mathbf{r}}_j$ onto the line-of-sight.

We reconstruct the pairwise kSZ signal in eight bins: 7 bins linearly separated between comoving pair separation $r$ of 40 and 200 Mpc, plus a final bin that includes pairs separated by 200 Mpc to 300 Mpc. The choice of the minimum separation is motivated by the fact that the pkSZ template is derived within the linear regime, limiting the modelling of the pairwise velocities between halos below $r \lesssim 40$ Mpc (due to, e.g., non-linearities and redshift space distortions) and because the photo-$z$ errors significantly suppress the signal and increase the statistical uncertainties. We choose to have a single bin at larger separations given that the pkSZ signal mostly arises from smaller and intermediate comoving separations, so for scales larger than 200 Mpc the signal is significantly smaller.

### B. CMB map filtering and temperature extraction

The next step is to extract the CMB temperature shift due to the kSZ effect at the location of the clusters. We apply a matched filter to the map, using prior knowledge of the spectral and angular dependence of the kSZ effect and other signals, to maximize the signal-to-noise on the pkSZ signal. The filter $\boldsymbol{\Psi}$ for $N_\nu$ different observed frequencies $\nu$ is constructed in Fourier space as,

$$\boldsymbol{\Psi}(\nu, \mathbf{k}) = \sigma_{\boldsymbol{\Psi}}^2 \mathbf{N}^{-1}(\nu, \mathbf{k}) \cdot \mathbf{S}_{\rm filt}(\nu, \mathbf{k}), \qquad (5)$$

where $\mathbf{k}$ is the Fourier mode, $\mathbf{N}^{-1}(\nu, \mathbf{k})$ is the inverse of the noise covariance matrix of the maps, $\mathbf{S}_{\rm filt}(\nu, \mathbf{k})$ is the expected signal vector in Fourier space, and $\sigma_{\boldsymbol{\Psi}}^2$ is the predicted variance of the filtered map. In this work, we assume the cluster emission follows a projected isothermal $\beta$-profile [49] with $\beta = 1$, written in cluster-centric coordinates as,

$$T(\theta) = T_0(1 + \theta^2/\theta_c^2)^{-1}, \qquad (6)$$

with $\theta_c$ being the angular core radius of the cluster, which is taken as $\theta_c = 0.5'$ throught this work. The pkSZ results were found to be insensitive to the choice of $\theta_c$ at higher noise levels [23], however we did not test this effect for the SPT-3G data because the significance of detection of the pkSZ signal will not increase in a significant manner in comparison with [23] due to the instrinsic limit that come from photometric redshift uncertainties, as shown in Section V B. The expected signal template in the matched filter is the convolution of this $\beta$-profile, the instrumental beam, and map filtering.

We can estimate the variance of the filtered map from,

$$\sigma_\Psi^2 = \left[ \int d^2\mathbf{k}\, \mathbf{S}_{\text{filt}}^\dagger(\nu, \mathbf{k}) \cdot \mathbf{N}^{-1}(\nu, \mathbf{k}) \cdot \mathbf{S}_{\text{filt}}(\nu, \mathbf{k}) \right]^{-1}. \quad (7)$$

We assume the noise is stationary, allowing the noise covariance matrix of the maps $\mathbf{N}(\nu, \mathbf{k})$ to be expressed as a symmetrical $N_\nu \times N_\nu$ matrix at each value of $\mathbf{k}$, where the diagonal elements are the auto-power spectra of every frequency map and the off-diagonal elements are the cross-spectra between the different frequencies.

The filter is built such that $\hat{T}_0$, an estimate of $T_0$, is extracted when centered on the cluster at position $\hat{\mathbf{n}}_0$ as

$$\hat{T}_0 = \int d^2\hat{\mathbf{n}}\, \boldsymbol{\Psi}(\nu, \hat{\mathbf{n}} - \hat{\mathbf{n}}_0) \cdot \mathbf{T}(\nu, \hat{\mathbf{n}}), \quad (8)$$

where $\mathbf{T}(\nu, \hat{\mathbf{n}})$ represents the vector of the temperature maps at different observed frequencies.

We use three different matched filters in this work for different purposes. The first filter is a minimum variance multi-frequency matched filter (MF-MF) with the 95, 150, and 220 GHz maps from SPT-3G. The second filter is a constrained minimum variance version (MF-tSZ), with the non-relativistic tSZ effect nulled, following [50]. Thirdly, following [23], we build a single-frequency matched filter for the 150 GHz map (MF-150GHz); the 150 GHz map has the lowest noise level of the three frequency bands at $3.9\,\mu$K-arcmin.

### C. Redshift-dependent foregrounds

Over an extended redshift range, the redshift evolution of tSZ signal and cosmic infrared background (CIB) emission can potentially introduce a redshift-dependent bias in the estimated temperatures. To mitigate any such redshift-dependent effects, we estimate the mean measured temperature as a function of redshift and subtract this mean temperature from the matched-filtered temperature values $\hat{T}_0(\hat{\mathbf{n}}_i)$, as

$$T(\hat{\mathbf{n}}_i) = \hat{T}_0(\hat{\mathbf{n}}_i) - \frac{\sum_j \hat{T}_0(\hat{\mathbf{n}}_j)\, G(z_i, z_j, \Sigma_z)}{\sum_j G(z_i, z_j, \Sigma_z)}. \quad (9)$$

The smoothed temperature at $z_i$ is calculated from the weighted sum of contributions of clusters at redshift $z_j$ using a Gaussian kernel $G(z_i, z_j, \Sigma_z) = \exp[-(z_i - z_j)^2/(2\Sigma_z^2)]$. For this analysis, we choose $\Sigma_z = 0.02$ resulting in a smooth temperature evolution. The choice of $\Sigma_z$ does not impact the result in any significant way [19, 23].

### D. Analytical modeling of the photo-$z$ uncertainties

Redshift uncertainties are the dominant source of error in the calculation of the separation distance between cluster pairs. The redshift errors, $\sigma_z$, leads to a rms uncertainty in the comoving distances, $\sigma_{d_c} = c\sigma_z/H(z)$ [23]. For the sample used in this work we find $\sigma_{d_c} \simeq 80$ Mpc. Redshift errors completely dilute the signal at $r \ll \sigma_{d_c}$, the signal is significantly reduced on scales $r \sim \sigma_{d_c}$, and the signal from cluster pairs with $r \gg \sigma_{d_c}$ is unaffected. Following the prescription from [23], we account for the smoothing due to the uncertain distances by multiplying the pairwise kSZ template in Eq. 2 by an exponential term to suppress the signal at small scales:

$$T_{\text{pkSZ}}(r, a) = \bar{\tau}_e \frac{v_{12}(r)}{c} T_{\text{CMB}} \times \left[1 - \exp\left(-\frac{r^2}{2\sigma_r^2}\right)\right]. \quad (10)$$

As in [23], we take the smoothing scale to be $\sigma_r = \sqrt{2}\sigma_{d_c}$. We test the analytic approach with simulations and find good agreement as shown in Fig. 3.

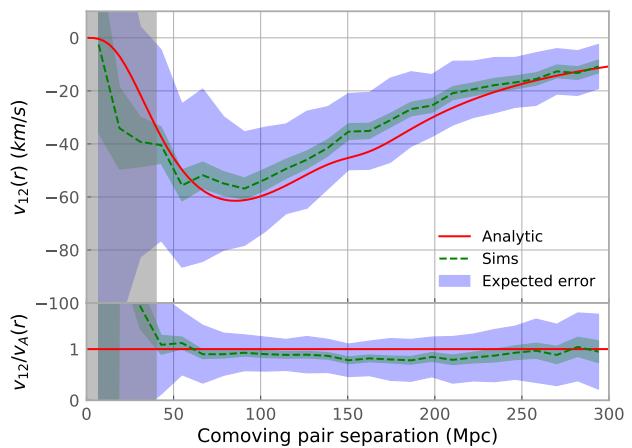

FIG. 3. A comparison between the analytical model (solid red line) of the mean pairwise velocity $v_{12}(r)$ compared to one obtained through simulations described in Section V A (dashed green line), corrected with Eq. 10 to account for the Gaussian photo-$z$ errors of the clusters of $\sigma_z \sim 0.01$, similar to the one measured for the DES cluster catalog. The wider shaded blue area shows the expected error bars from the pkSZ reconstruction for 2 years of SPT-3G temperature maps. The simulated cluster sample contains N=22,923 clusters within a mass range of $0.6 < M_{500c}/10^{14} M_\odot < 4$ between redshifts of $0.1 < z < 0.8$. The grey shaded region indicates separations $r < 40$ Mpc, where the analytical model breaks due to the non-linear regime.

### E. Covariance matrix

We estimate the covariance matrix of the binned pkSZ measurement directly from the data using two resampling techniques.

- **Jackknife**: The jackknife resampling technique (labelled 'JK' in equations) consists of measuring

the pkSZ signal by splitting the cluster catalogue into $N_{\rm JK}$ subsamples, removing one of them, and recomputing the pkSZ amplitude from the remaining $N_{\rm JK} - 1$ subsamples. This process is repeated until every subsample has been discarded once from the measurement. Then we estimate the covariance matrix as

$$\hat{C}_{ij}^{\rm JK} = \frac{N_{\rm JK} - 1}{N_{\rm JK}} \sum_{\alpha=1}^{N_{\rm JK}} (\hat{T}_i^\alpha - \bar{T}_i)(\hat{T}_j^\alpha - \bar{T}_j), \qquad (11)$$

where $\hat{T}_i^\alpha$ is the pairwise kSZ signal in separation bin $i$ and jackknife realization $\alpha$, of mean $\bar{T}_i$. For our main analysis, we use $N_{\rm JK} = 1,000$ subsamples.

- **Bootstrap**: The bootstrap method (indicated by 'BS' in equations) consists of randomly drawing with replacement an equal number of clusters, and recomputing the pkSZ signal for each random draw. This process is repeated $N_{\rm BS}$ times, and the covariance matrix $\hat{C}_{ij}^{\rm BS}$ estimated as:

$$\hat{C}_{ij}^{\rm BS} = \frac{1}{N_{\rm BS} - 1} \sum_{\alpha=1}^{N_{\rm BS}} (\hat{T}_i^\alpha - \bar{T}_i)(\hat{T}_j^\alpha - \bar{T}_j). \qquad (12)$$

Here $i$ and $j$ refer to the separation bin, $\hat{T}_i^\alpha$ is the estimated pkSZ signal in bin $i$ for the $\alpha$ random sample, and $\bar{T}_i$ the average pkSZ value across all samples. The bootstrap method is expected to need more random samples than the jackknife method to converge due to random sampling. As a result, it is more computationally expensive. We use $N_{\rm BS} = 10,000$ samples when reporting results with the bootstrap covariance.

The baseline covariance matrix in this work is estimated using the jackknife subsampling technique with 1,000 subsamples, and the correlation matrix derived from it is shown in Fig. 4. As a test of robustness, we also show selected results when the covariance is estimated from a different number of subsamples or the bootstrap technique. We show in Fig. 5 a comparison of the pkSZ error bars calculated from the two methods. Both estimators have clearly converged and show minimal differences in the covariance values between $N_{\rm JK} = 1,000$ or 2,000, and $N_{\rm BS} = 4,000$ or 10,000. The bootstrap estimator yields larger uncertainties at small separation distances, however the differences are within allowable tolerances for the current signal-to-noise.

For the inverse of the covariance, we use the estimator

$$\tilde{C}^{-1} = \frac{N - N_{\rm bins} - 2}{N - 1} \hat{C}^{-1} \qquad (13)$$

where $N$ is the number of jackknife or bootstrap samples used to compute the covariance matrix $\hat{C}^{\rm JK/BS}$, and $N_{\rm bins}$ is the number of comoving separation bins. This correction factor is needed because the empirically determined inverse covariance matrix $\hat{C}^{-1}$ is a biased estimator of the true inverse covariance matrix $C^{-1}$ as shown in [51].

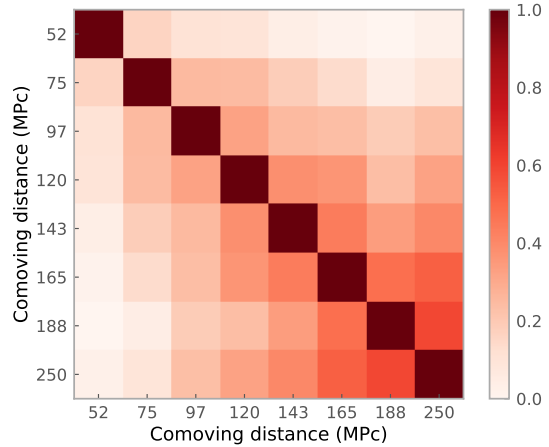

FIG. 4. Correlation matrix of the pkSZ measurement shown in Fig. 6 calculated with 1,000 jackknife subsamples. The higher distance bins show more correlation because on average we encounter the same clusters more times.

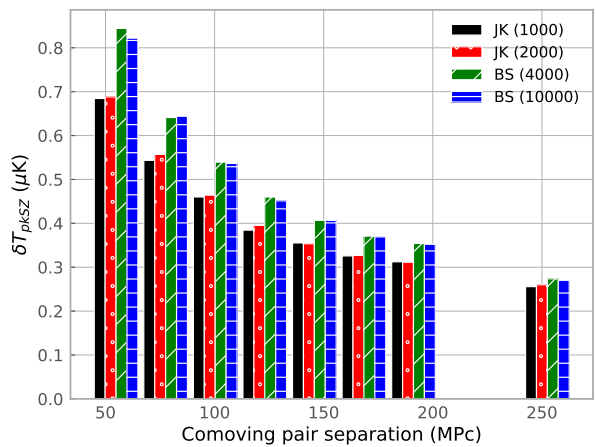

FIG. 5. Estimated uncertainties from jackknife (JK) and bootstrap (BS) covariance estimators for 1000/2000 subsamples and 4000/10000 resamplings, respectively. The error estimate is stable across the different methods and number of subsamples/resamples. We construct this figure from the SPT-3G + DES-Y3 data.

### F. Model fitting and statistical significance

We fit the measured pkSZ signal to a one parameter model, scaling the analytical template given by Eq. 10 by the unknown average optical depth of the cluster sample

$\bar{\tau}_e$. We then compute the statistical significance of our measurement in two different ways:

1. The main results will be presented by obtaining the best-fit $\bar{\tau}_e$ and its uncertainty by minimizing the $\chi^2$ as

$$\chi^2(\bar{\tau}_e) = [\hat{T}_{\text{pkSZ}} - T_{\text{pkSZ}}(\bar{\tau}_e)]^\dagger \tilde{C}^{-1} [\hat{T}_{\text{pkSZ}} - T_{\text{pkSZ}}(\bar{\tau}_e)]. \quad (14)$$

The signal-to-noise ratio $S/N$ is then computed with $S/N = \bar{\tau}_e/\sigma_{\bar{\tau}_e}$, where $\sigma_{\bar{\tau}_e}$ is given by $\chi^2(\bar{\tau}_e \pm \sigma_{\bar{\tau}_e}) - \chi^2_{\min} = 1$.

2. To complement the previous significance, we also assess the signal significance by calculating the $\chi^2$ with respect to the null-signal hypothesis:

$$\chi^2_0 = \hat{T}^\dagger_{\text{pkSZ}} \tilde{C}^{-1} \hat{T}_{\text{pkSZ}}. \quad (15)$$

We estimate the probability-to-exceed the observed $\chi^2_0$ (PTE) by comparing it to the cumulative distribution function (CDF) of the $\chi^2$ distribution. The PTE provides one estimate for how likely it is that the data could result from a noisy measurement of zero pkSZ signal.

We expect the template fit to yield a higher statistical significance than the null-signal procedure due to the fact that the first one includes the additional information of our analytic template, whereas the latter one makes no assumptions about the expected signal shape.

## V. SIMULATIONS

### A. Simulations of the mm-wave sky

In order to validate and test the accuracy of our analysis pipeline, as well as to estimate the impact of systematic effects, we use realistic realizations of the millimeter wavelength sky from the MDPL2 Synthetic Skies suite [Omori, in prep.]. The simulated skies are generated by pasting astrophysical effects onto the halo lightcone from the MultiDark Planck 2 $N$-body simulation [52]. The astrophysical modeling in the simulation has been calibrated using observational data and external hydrodynamical simulations. Outlined below are the main components of the simulated microwave sky.

- The dark matter density field is used to gravitationally lens the CMB sky.

- The tSZ signal from each dark matter halo is added based on the [53] electron thermal pressure profile that was calibrated on the hydrodynamical BAHAMAS simulations suite [54].

- The kSZ effect is added in a similar way. The same [53] gas profile is used to estimate the electron number density, which is multiplied by the line-of-sight velocity to obtain the kSZ signal from the halo.

- The cosmic infrared background (CIB) from dust-enshrouded galaxies is simulated by first assigning star formation rate and stellar mass to each individual halo using the UniverseMachine code [55]. With that information, the bolometric infrared luminosity is inferred from [56], and then converted to flux density assuming the shape of the spectral energy distribution to be a modified blackbody.

- We add instrumental noise to the simulations, assuming white noise levels of 7, 5, and 20 $\mu$K-arcmin at 90, 150, and 220 GHz, respectively. These noise levels are similar to the real data maps at the angular scales of interest, $3000 < \ell < 5000$.

The mock cluster catalog is generated by first mapping the redMaPPer clusters richness $\lambda$ to the mass within a spherical region with an average density of 500 times the critical density $M_{500c}$ according to the weak lensing mass calibration from [57], and then by selecting the objects in the MDPL2 halo catalogue that lie in this mass range. Within the SPT-3G footprint, the simulated cluster sample contains N=22,923 clusters within a mass range of $0.6 < M_{500c}/10^{14} M_\odot < 4$, which is similar to the $10 < \tilde{\lambda} < 60$ richness range from the DES cluster catalog that we will use in our real dataset.

### B. Pipeline validation

We now use the simulation suite introduced above to explore the sensitivity of the pairwise kSZ estimator to the presence of contaminating signals and the noise level in the data set.

We check how the instrumental noise level of the 150 GHz map affects the recovered pkSZ signal, as well as the impact of photometric redshift uncertainties $\sigma_z$ on the measured kSZ amplitude and report our findings in Table I. We extract the temperature at the clusters' positions for this test using the single frequency matched filter and use the same range of scales as the one adopted for the real analysis described in Section IV A. As expected, we find that decreasing the instrumental noise levels translates to an increased detection significance; however, when we include the redshift errors ($\sigma_z > 0$) the significance level does not improve. This indicates that redshift uncertainties pose an intrinsic limit to this analysis. We also choose a noise level of 18 $\mu$K-arcmin to approximate the noise level from the SPT-SZ CMB maps used in the analysis of [23]. They estimated a signal from a different set of simulations [58] at $3.7\sigma$, with a mass range of $0.9 < M_{500c}/(10^{14} M_\odot) < 4$. We obtain within the same mass range a $S/N$ of $3.8\sigma$, agreeing with their estimates and giving us confidence in the simulations that we are using. The 5 $\mu$K-arcmin corresponds approximately to the current noise levels of the 150 GHz map with one year of the SPT-3G data. From now on, we will use $\sigma_z = 0.01$ since it is approximately the root mean square photo-$z$ error for the DES cluster catalogue (see Section III B).

| Noise level ($\mu$K-arcmin) | S/N ($\sigma_z = 0$) | S/N ($\sigma_z = 0.01$) |
|---|---|---|
| 18 | 7.8 | 3.8 |
| 5 | 10.4 | 3.9 |

TABLE I. Impact of the 150 GHz map instrumental noise levels on the $S/N$ of pkSZ. We see how lowering the map noise level increases the detection significance of the signal, but the photometric redshift error $\sigma_z$ dominates the signal. These results are obtained by including all the cosmological and foreground components (CMB, tSZ, CIB) with the noise levels noted on the first column.

| Foreground removed | S/N |
|---|---|
| tSZ | 4.8 |
| CMB | 5.3 |
| CIB | 4.3 |

TABLE II. Effect of different foreground removals on the 150 GHz map on the $S/N$ of pkSZ, where all the results are higher than the $3.9\,\sigma$ found in Tab. I. For these results we fix the noise levels to $5\mu$K-arcmin and include photo-$z$ errors with $\sigma_z = 0.01$.

| Method | Simple sample | Mixed sample |
|---|---|---|
| MF-MF | 3.6 | 3.4 |
| MF-tSZ | 4.0 | 3.8 |
| MF-tSZ-CIB | 3.6 | 3.8 |
| MF-MF-CIB | 3.2 | 3.2 |

TABLE III. Effect on the pkSZ $S/N$ of the CIB subtraction by increasing its power in the covariance matrix for the matched filter construction. MF-MF stands for matched filter multi-frequency, MF-tSZ stands for matched filter with a tSZ deprojection, and -CIB stands for a CIB reduction following [59] for each of the previous matched filters. The simple and mixed sampled are described in the last paragraph of Section V B.

We also investigate the impact of different foregrounds on the statistical uncertainties by running the extraction pipeline on maps that have primary CMB, tSZ, and CIB set to zero one by one, while keeping the rest of the foregrounds unchanged for the 150 GHz map. Removing the foregrounds clearly helps to improve the signal as shown in Table II. In particular, the removal of tSZ results in an increase in the $S/N$ ratio by 25%. To remove the contamination from cluster tSZ signal, we explore the use of multifrequency matched filters (MF-MF), including a version in which the particular frequency dependence of the tSZ is used to deproject it explicitly (MF-tSZ). This deprojection, however, comes with a noise penalty that may be larger than the tSZ contamination itself, particularly for lower-mass clusters.

Finally, we also artificially increase the CIB signal power by a factor of 5 in the noise covariance used for the multi-frequency matched filter with and without tSZ deprojection (MF-tSZ-CIB and MF-MF-CIB, respectively) in an attempt to reduce its effect as shown in [59]. We tested a mix (Mixed Sample) of the cluster signal where the low-mass clusters ($0.6 < M_{500c}/10^{14}M_\odot < 1$) are extracted from the single frequency matched filtered 150 GHz map (N=14,321), while the high mass clusters' ($1 < M_{500c}/10^{14}M_\odot < 4$) temperatures are extracted from the tSZ deprojected multifrequency matched filtered map (N=8,602). We obtain similar significance levels with all the different matched filters and cluster samples, with results shown in Table III. Since the significance levels are similar, we will not try to suppress the CIB in the SPT-3G data.

## VI. PAIRWISE KSZ MEASUREMENT

### A. Pairwise kSZ signal from SPT and DES

The pairwise kSZ measurement from SPT-3G maps and the full DES Year-3 redMaPPer cluster catalog in the $10 < \tilde{\lambda} < 60$ richness range ($N = 24,580$) is presented in Fig. 6. This result has been obtained by combining the temperatures extracted from the tSZ deprojected map for high richness clusters ($30 < \tilde{\lambda} < 60$) and temperatures extracted from the matched filtered 150 GHz map for lower richness clusters ($\tilde{\lambda} < 30$). As discussed in Section IV F, we estimate and report a detection at a significance of $4.1\,\sigma$. The result of the fit to the analytical pairwise kSZ template yields a cluster catalog mean optical depth of

$$\bar{\tau}_e = (2.97 \pm 0.73) \times 10^{-3}. \tag{16}$$

The corresponding correlation matrix between different radial separations is shown in Fig. 4. [23] found $\bar{\tau}_e = (1.37 \pm 0.41) \times 10^{-3}$ for the same richness range and 28,760 clusters using SPT-SZ and the DES-Y1 cluster catalog, which is $\sim 2\,\sigma$ lower than the value we find. Another previous analysis [24] found $\bar{\tau}_e = (0.69 \pm 0.34) \times 10^{-4}$ for a mass range of $1 < M_{200c}/10^{13}M_\odot < 1.6$. The average mass of these clusters is an order of magnitude less massive than the estimated mass of our catalog $0.6 < M_{500c}/10^{14}M_\odot < 4$, thus finding a higher value of the $\bar{\tau}_e$ in our analysis is consistent.

As noted earlier, the total significance of the pkSZ detection in this work ($4.1\,\sigma$) is similar to that from [23] ($4.2\sigma$), despite a large improvement in CMB map noise. As discussed in Sec. V B, this is because the redshift uncertainties pose an intrinsic limit to the analysis.

As a consistency check, we compare the detection significances obtained using the alternative matched filters introduced in Section IV B. We first apply these matched filters to the SPT-3G maps to extract the temperature at the clusters' positions and then we reconstruct the pairwise kSZ for each of them. The results of the fits to the analytical pkSZ template are displayed in Tab. IV. As can be seen, all the $\bar{\tau}_e$ values are well within the $1\sigma$ statistical uncertainties of each other and the corresponding detections shift by less than $0.5\sigma$.

Finally, we explore how the detection is affected by a higher low-mass threshold by repeating the analysis for the richness range $20 < \tilde{\lambda} < 60$. The results of this test are reported in the right column of Tab. IV, where we can clearly see that the significance of the detection has decreased greatly due to the limited number of clusters that fall in this richness range, thus increasing the estimated errors on the measurements.

| Method | $10 < \tilde{\lambda} < 60$ ($N = 24,580$) | $20 < \tilde{\lambda} < 60$ ($N = 5,797$) |
|---|---|---|
| MF-150GHz | $3.08 \pm 0.75$ (4.1) | $2.39 \pm 1.65$ (1.4) |
| MF-MF | $2.85 \pm 0.89$ (3.2) | $2.16 \pm 2.13$ (1.0) |
| MF-tSZ | $3.72 \pm 1.15$ (3.2) | $2.61 \pm 2.03$ (1.3) |
| Mixed | **$2.97 \pm 0.73$ (4.1)** | $2.66 \pm 1.65$ (1.6) |

TABLE IV. The mean optical depth $\bar{\tau}_e \times 10^3$ and the $S/N$ of each one in parenthesis, for two main richness cuts taken on the DES catalog for this analysis. The different methods to extract the temperature at the clusters' positions are explained in Section V B. MF-150GHz stands for a matched filter for only the 150GHz temperature map, MF-MF stands for matched filter multifrequency, MF-tSZ stands for matched filter with a tSZ deprojection, and Mixed stands for the mixed catalog of low mass clusters coming from the MF-150GHz and higher mass clusters coming from the MF-tSZ. The baseline result of the paper is highlighted in bold.

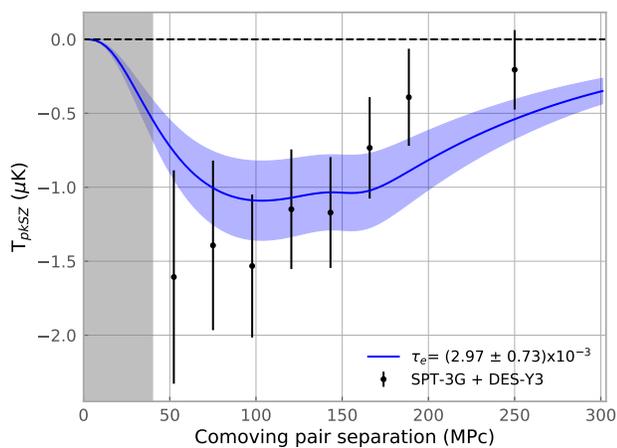

FIG. 6. We detect the pkSZ signal at $4.1\sigma$, using the covariance estimated with the jackknife method and 1,000 subsamples. As described in Section IV E, the 150 GHz SPT-3G map is used for low-richness clusters ($\lambda \leq 30$) while a multifrequency matched filter tSZ-free map is used for high-richness clusters ($\lambda > 30$). The recovered mean optical depth of the cluster sample is $\bar{\tau}_e = (2.97 \pm 0.73) \times 10^{-3}$. The grey shaded region indicates the scales ($r < 40$) Mpc where the analytical model breaks due to the non-linear regime.

### B. Null tests

We run a suite of null tests that check whether the signal present in the data has statistical properties consistent with the pairwise kSZ effect

- **Sign-flip**: For this test, we replace the minus sign inside the sum in the estimator in Eq. 4 with a plus sign to remove sensitivity to the pkSZ signal.

- **Position-shuffling**: By randomly shuffling the redshifts of the clusters while keeping their extracted temperatures unchanged, we null the pairwise signal by making $c_{ij}$ maximum on clusters that are not under the gravitational influence of each other.

- **Temperature-shuffling**: We randomly shuffle the clusters' extracted temperature without changing their position, keeping the same $c_{ij}$ for the estimator and thus removing the pairwise signal from the clusters.

As shown in Fig. 7, the null tests remove the pairwise kSZ signal, leaving a mean-zero signal with correlated uncertainties as encoded in the covariance matrix. To quantify the result of these tests, we calculate the reduced $\chi^2$ for each bootstrap resample in each test, and we quote the probability-to-exceed (PTE) as the fraction of bootstrap resamples with reduced $\chi^2 > 1$. We obtain PTEs of 76%, 48% and 62% for the sign-flip, distance shuffling and temperature shuffling respectively, with a mean reduced $\chi^2 \sim 1$ for each of the tests. These null tests are consistent with no detection, giving us confidence in our measurement of the pairwise kSZ effect.

We also use the null test bootstrap resamples as a check of our formal estimate of the uncertainty on $\bar{\tau}_e$. We display the distribution of bootstrap resamples for the sign-flip test in Fig. 8; the distributions for the other tests look similar. The estimated error of the mean optical depth from real data is comparable with our null tests errors within $\sim 20\%$, which gives us confidence on the accuracy of the measurement.

### C. Systematics tests

We test some systematics that could influence our measurement of the pkSZ signal in order to quantify any im-pact on our analysis and subsequent measurements.

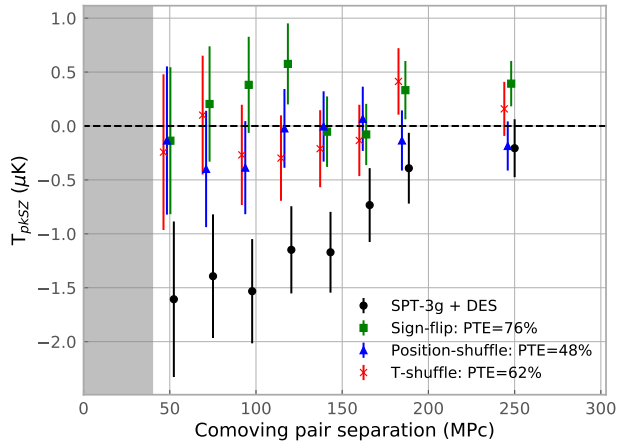

FIG. 7. Null tests for the SPT-3G + Full DES catalog pkSZ measurements. All the null-tests yield a reduced $\chi^2 \sim 1$ and their PTE values are reported in the legend. The black points are the data points measured of the pkSZ signal for our baseline analysis. The points are offset in comoving separation for visualization purposes.

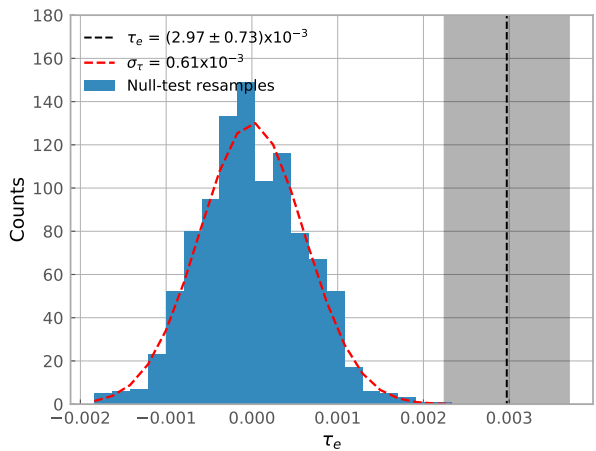

FIG. 8. Histogram of 1,000 bootstrap resamples for the sign-flip test estimating $\bar{\tau}_e$, comparing them to the baseline result of $\bar{\tau}_e$ shown in the dashed black line, with the grey region representing the $1\sigma$ uncertainty. This histogram shows us how a resampling of null tests produces an estimated Gaussian error for the $\bar{\tau}_e$ of $\sigma_{\bar{\tau}_e} = 0.61 \times 10^{-3}$, which is a ~20% difference to the one obtained for the real data of $\sigma_{\bar{\tau}_e} = 0.73 \times 10^{-3}$.

- **Mass scatter**: In order to match the mass range from simulations, where the masses of clusters are known, to the optical cluster catalog that is selected in richness, we need a good understanding of how to obtain an accurate representation of the mass ranges under analysis. This is of particular interest because the analytical model in Eq. 3, which is used to infer the clusters' optical depth, depends on the mass range of interest and changing the typical cluster mass could significantly bias this result. In this work we have selected the simulation sample using the relation shown in [57]. However, we need to take into account that these are estimated measurements, therefore a scatter in the cluster mass of the optical data can occur. To model this scatter, we draw mass errors from a normal distribution with width $\sigma_{\ln(M)} = 0.3$, which is an underestimation of the scatter at low richness for a Gaussian error model given the significant projection effects in the DES sample [45, 60–62], but it gives an idea of how significant the mass scattering can be. We then use these errors to compute the pkSZ signal from the simulations, obtaining an average decrease on the signal detection to $2.5\sigma$ on the simulation catalog for the mass range of $0.6 < M_{500c}/10^{14} M_\odot < 4$. This implies that the measured pkSZ significance might be $\sim 1\sigma$ lower than it could be due to this effect.

- **Mis-centering**: The measured pkSZ signal can be diluted due to the fact that the clusters' positions estimated from the optical survey catalog might not coincide with the location of the cluster kSZ signal. This mis-centering has a larger impact on clusters that are not fully relaxed or are merging, where the potential minimum is not located on the brightest cluster galaxy, or where this galaxy has been misidentified by the redMaPPer algorithm. The impact of mis-centering has been tested before [23], where two different mis-centering models [46, 63] were tested and identified a reduction of $\sim 10\%$ of the pkSZ significance. Although for our confidence levels it does not produce a significant impact, it should be considered for future spectroscopic redshift catalogues.

### D. Estimating the optical depth from the thermal SZ effect

In addition to the bulk velocity of electrons, the electrons' random thermal motion imprints a signature on the observed CMB through inverse-Compton scattering, the thermal Sunyaev-Zel'dovich (tSZ) effect [e.g., 4, 64]. The magnitude of the tSZ effect produced along a line-of-sight $\hat{\mathbf{n}}_i$ can be quantified by the Compton $y$-parameter [65],

$$y(\hat{\mathbf{n}}_i) = \int d\ell \, n_e \frac{k_B T_e}{m_e c^2} \sigma_T, \qquad (17)$$

where $k_B$ is the Boltzmann constant, $m_e$ is the mass of the electron, and $T_e$ is the electron temperature. The tSZ effect induces a frequency-dependent shift of the observed CMB temperature, which in the non-relativistic limit can

be written as:

$$\frac{\Delta T}{T_{\text{CMB}}} = g(\nu)y, \quad (18)$$

with $g(\nu) = x\frac{e^x+1}{e^x-1} - 4$, and $x = h\nu/(k_B T_{\text{CMB}})$. The frequency dependence of the tSZ effect is such that the effect appears as a temperature decrement at lower frequencies than $\sim 218$ GHz, while being completely null at that frequency value [4]. Since the tSZ is directly related to the electron pressure (number density of electrons times the electron temperature) of the cluster, the signal becomes stronger for more massive clusters.

We build a $y$-map by performing an internal linear combination (ILC) [66] on the 90 GHz, 150 GHz, and 220 GHz temperature maps from SPT-3G. This $y$ map is preliminary and has not been fully optimized, but we are primarily concerned with the mean value of cluster optical depth, and a non-optimized $y$-map will mainly result in elevated variance, and not bias, in the optical depth measurement. Using this $y$-map, we stack all our clusters and extract the average $y$ value through aperture photometry. The aperture photometry filter is written in real-space as

$$\Psi(\theta) = \begin{cases} 1 & 0 < \theta < \theta_r \\ -1 & \theta_r < \theta < \sqrt{2}\theta_r \\ 0 & \text{elsewhere} \end{cases}, \quad (19)$$

where $\theta_r$ is the characteristic filter scale. The aperture photometry effectively reduces the noise on all scales that are larger than the filter scale by subtracting the average temperature in the outer ring from the average temperature inside the disc of radius $\theta_r$. In contrast to the matched filter technique described in Section IV B, this approach does not assume a specific model for the cluster profile, however it requires that the cluster is contained within the characteristic filter scale $\theta_r$ to avoid biases in the temperature estimation.

We follow [28] and relate the mean $y$-value of the clusters to the mean optical depth $\bar{\tau}_e$ according to:

$$\ln(\bar{\tau}_e) = \ln(\tau_0) + m\ln(\bar{y}). \quad (20)$$

For the signal-to-noise-maximizing filter scale of $\theta_r = 2.6'$, [28] calibrated the coefficients to be $\ln(\tau_0) = -6.47$ and $m = 0.49$.

The transfer function and beam applied to the SPT-3G temperature maps described in Section III produce a bias for object-based analysis such as aperture photometry [59]. To estimate this bias, we applied the same transfer function and beam filters to the set of simulations in Section V A. We compute the aperture photometry with $\theta_r = 2.6'$ to the recovered filtered $y$-map and the original one. We then compute the $\bar{\tau}_e$ finding a 10% reduction of the measured filtered value in comparison with the original simulation $y$-map. Taking this into account, we find

$$\bar{\tau}_e = (2.51 \pm 0.55^{\text{stat}} \pm 0.15^{\text{syst}}) \times 10^{-3}, \quad (21)$$

where we estimate the statistical (stat) uncertainty of this measurement by performing 1,000 JK resamples to the $y$-values of the clusters, while the systematic errors are obtained by propagating the uncertainty from the calibrated values (syst) $\ln(\tau_0)$ and $m$, which are 2% and 6% respectively.

The Compton-$y$ based estimate of the mean optical depth is within $0.6\,\sigma$ of the pkSZ-derived estimate of $\bar{\tau}_e = (2.97 \pm 0.73) \times 10^{-3}$. Future works will be able to use the Compton-$y$ estimate of the mean optical depth to break the degeneracy between the optical depth and velocity in the pkSZ signal, and significantly improve tests of cosmology from the pkSZ effect.

## VII. CONCLUSIONS

In this work, we measure the mean optical depth of the DES-Y3 redMaPPer cluster sample in the $10 < \tilde{\lambda} < 60$ richness range and between $0.1 < z < 0.8$ and find the best-fit value to be $\bar{\tau}_e = (2.97 \pm 0.73) \times 10^{-3}$. The optical depth measurement is derived from a $4.1\,\sigma$ detection of the pairwise kSZ effect. The SPT-3G and DES surveys overlap over $\sim 1,400$ deg$^2$ of southern sky, and after cuts, there are 24,580 galaxy clusters from the DES-Y3 redMaPPer cluster sample within the SPT-3G survey region. We extract the CMB temperature shift at the location of these clusters using a matched filter approach to optimize signal-to-noise in the $150\,\text{GHz}$ maps for the low-mass end of the cluster sample ($10 < \tilde{\lambda} < 30$), and a constrained matched filter to zero the non-relativistic tSZ effect for more massive clusters ($30 < \tilde{\lambda} < 60$).

We validate the analysis using simulated data from the MDPL2 simulation suite [Omori, in prep.]. We also use these simulations to explore the limiting uncertainties in the analysis, finding the major sources of uncertainty in the current data set to be due to uncertain cluster redshifts and, if not removed, the tSZ effect in massive clusters. This result motivates the decision to use a constrained matched filter to zero the thermal SZ signal in clusters with a richness $\tilde{\lambda} > 30$. There are also non-negligible contributions from the primary CMB anisotropy, cosmic infrared background, and instrumental noise.

We test the robustness of the detection by repeating the analysis using different methods of temperature extraction, finding agreement between the recovered optical depth and the $S/N$ levels. We also found an agreement on optical depth when increasing the minimum richness ($20 < \tilde{\lambda} < 60$), although with a lower significance on the pkSZ signal due to fewer clusters in this cut. To provide further evidence of the robustness of our results, we have conducted different null-tests where we artificially remove any cosmological signal and found the recovered pkSZ measurement to be consistent with zero.

Finally, we compare our result for the mean optical depth of the cluster sample $\bar{\tau}_e$ from the pkSZ measurement to one obtained based on the mean Compton $y$- parameter as described in [28], finding the two estimates agree within $0.6\,\sigma$. This demonstrates the application of using the observed thermal SZ signal to break the degeneracy between the mean optical depth and velocity for the pkSZ effect. The combination of upcoming CMB and spectroscopic surveys is expected to yield high significance measurements of the pkSZ signal.

By breaking the degeneracy with astrophysics using alternative techniques like this, we can proceed to constrain cosmological parameters using the pkSZ; however, for this to occur a higher signal-to-noise of the pkSZ signal is required. Assuming current SPT-3G CMB map noise levels, we expect that future spectroscopic catalogues will significantly reduce clusters' redshift uncertainty, leading to an increase of the pkSZ signal to $\sim 10\sigma$. This will increase further in future CMB experiments with higher sky coverage and lower noise levels.

## ACKNOWLEDGMENTS

SPT is supported by the National Science Foundation through grants PLR-1248097 and OPP-1852617. Partial support is also provided by the NSF Physics Frontier Center grant PHY-1125897 to the Kavli Institute of Cosmological Physics at the University of Chicago, the Kavli Foundation and the Gordon and Betty Moore Foundation grant GBMF 947. This research used resources of the National Energy Research Scientific Computing Center (NERSC), a DOE Office of Science User Facility supported by the Office of Science of the U.S. Department of Energy under Contract No. DE-AC02-05CH11231. The Melbourne group acknowledges support from the Australian Research Council's Discovery Projects scheme (DP200101068). B.B. is supported by the Fermi Research Alliance LLC under contract no. De-AC02- 07CH11359 with the U.S. Department of Energy. Argonne National Laboratory's work was supported under the U.S. Department of Energy contract DE-AC02-06CH11357. We also acknowledge support from the Argonne Center for Nanoscale Materials.

This research was done using services provided by the OSG Consortium [67, 68], which is supported by the National Science Foundation awards #2030508 and #1836650.

Funding for the DES Projects has been provided by the U.S. Department of Energy, the U.S. National Science Foundation, the Ministry of Science and Education of Spain, the Science and Technology Facilities Council of the United Kingdom, the Higher Education Funding Council for England, the National Center for Supercomputing Applications at the University of Illinois at Urbana-Champaign, the Kavli Institute of Cosmological Physics at the University of Chicago, the Center for Cosmology and Astro-Particle Physics at the Ohio State University, the Mitchell Institute for Fundamental Physics and Astronomy at Texas A&M University, Financiadora de Estudos e Projetos, Fundação Carlos Chagas Filho de Amparo à Pesquisa do Estado do Rio de Janeiro, Conselho Nacional de Desenvolvimento Científico e Tecnológico and the Ministério da Ciência, Tecnologia e Inovação, the Deutsche Forschungsgemeinschaft and the Collaborating Institutions in the Dark Energy Survey.

The Collaborating Institutions are Argonne National Laboratory, the University of California at Santa Cruz, the University of Cambridge, Centro de Investigaciones Energéticas, Medioambientales y Tecnológicas-Madrid, the University of Chicago, University College London, the DES-Brazil Consortium, the University of Edinburgh, the Eidgenössische Technische Hochschule (ETH) Zürich, Fermi National Accelerator Laboratory, the University of Illinois at Urbana-Champaign, the Institut de Ciències de l'Espai (IEEC/CSIC), the Institut de Física d'Altes Energies, Lawrence Berkeley National Laboratory, the Ludwig-Maximilians Universität München and the associated Excellence Cluster Universe, the University of Michigan, NFS's NOIRLab, the University of Nottingham, The Ohio State University, the University of Pennsylvania, the University of Portsmouth, SLAC National Accelerator Laboratory, Stanford University, the University of Sussex, Texas A&M University, and the OzDES Membership Consortium.

Based in part on observations at Cerro Tololo Inter-American Observatory at NSF's NOIRLab (NOIRLab Prop. ID 2012B-0001; PI: J. Frieman), which is managed by the Association of Universities for Research in Astronomy (AURA) under a cooperative agreement with the National Science Foundation.

The DES data management system is supported by the National Science Foundation under Grant Numbers AST-1138766 and AST-1536171. The DES participants from Spanish institutions are partially supported by MICINN under grants ESP2017-89838, PGC2018-094773, PGC2018-102021, SEV-2016-0588, SEV-2016-0597, and MDM-2015-0509, some of which include ERDF funds from the European Union. IFAE is partially funded by the CERCA program of the Generalitat de Catalunya. Research leading to these results has received funding from the European Research Council under the European Union's Seventh Framework Program (FP7/2007-2013) including ERC grant agreements 240672, 291329, and 306478. We acknowledge support from the Brazilian Instituto Nacional de Ciência e Tecnologia (INCT) do e-Universo (CNPq grant 465376/2014-2). This manuscript has been authored by Fermi Research Alliance, LLC under Contract No. DE-AC02-07CH11359 with the U.S. Department of Energy, Office of Science, Office of High Energy Physics.

The CosmoSim database used in this paper is a service by the Leibniz-Institute for Astrophysics Potsdam (AIP). The MultiDark database was developed in cooperation with the Spanish MultiDark Consolider Project CSD2009-00064. The authors gratefully acknowledge the Gauss Centre for Supercomputing e.V. (www.gauss-centre.eu) and the Partnership for Advanced Supercomputing in Europe (PRACE, www.prace-ri.eu) for funding the MultiDark simulation project by providing computing time on the GCS Supercomputer SuperMUC at Leibniz Supercomputing Centre (LRZ, www.lrz.de). The Bolshoi simulations have been performed within the Bolshoi project of the University of California High-Performance AstroComputing Center (UC-HiPACC) and were run at the NASA Ames Research Center.

We acknowledge the use of many python packages: IPYTHON [69], MATPLOTLIB [70], AND SCIPY [71].

[1] R. A. Sunyaev and Y. B. Zel'dovich, Ap&SS **7**, 3 (1970).
[2] R. Sunyaev and Y. Zel'dovich, ARAA **18**, 537 (1980).
[3] M. Birkinshaw, Physics Reports **310**, 97 (1999).
[4] J. E. Carlstrom, G. P. Holder, and E. D. Reese, ARA&A **40**, 643 (2002).
[5] S. Das, T. Louis, M. R. Nolta, G. E. Addison, E. S. Battistelli, J. R. Bond, E. Calabrese, D. Crichton, M. J. Devlin, S. Dicker, et al., J. of Cosm. & Astropart. Phys. **4**, 014 (2014), arXiv:1301.1037 [astro-ph.CO].
[6] C. L. Reichardt, S. Patil, P. A. R. Ade, A. J. Anderson, J. E. Austermann, J. S. Avva, E. Baxter, J. A. Beall, A. N. Bender, B. A. Benson, et al., arXiv e-prints , arXiv:2002.06197 (2020), arXiv:2002.06197 [astro-ph.CO].
[7] T. M. Crawford, K. K. Schaffer, S. Bhattacharya, K. A. Aird, B. A. Benson, L. E. Bleem, J. E. Carlstrom, C. L. Chang, H.-M. Cho, A. T. Crites, et al., Astrophys. J. **784**, 143 (2014), arXiv:1303.3535.
[8] W. R. Coulton, S. Aiola, N. Battaglia, E. Calabrese, S. K. Choi, M. J. Devlin, P. A. Gallardo, J. C. Hill, A. D. Hincks, J. Hubmayr, et al., Journal of Cosmology and Astroparticle Physics **2018** (09), 022.
[9] N. Huang, L. E. Bleem, B. Stalder, P. A. R. Ade, S. W. Allen, A. J. Anderson, J. E. Austermann, J. S. Avva, J. A. Beall, A. N. Bender, et al., arXiv e-prints , arXiv:1907.09621 (2019), arXiv:1907.09621 [astro-ph.CO].
[10] M. Hilton, C. Sifón, S. Naess, M. Madhavacheril, M. Oguri, E. Rozo, E. Rykoff, T. M. C. Abbott, S. Adhikari, M. Aguena, et al., The Astrophysical Journal Supplement Series **253**, 3 (2021).
[11] L. D. Shaw, D. H. Rudd, and D. Nagai, Astrophys. J. **756**, 15 (2012), arXiv:1109.0553 [astro-ph.CO].
[12] O. Lahav, M. J. Rees, P. B. Lilje, and J. R. Primack, MNRAS **251**, 128 (1991).
[13] M. G. Haehnelt and M. Tegmark, MNRAS **279**, 545+ (1996).
[14] Y.-Z. Ma and G.-B. Zhao, Phys. Lett. B **735**, 402 (2014), arXiv:1309.1163 [astro-ph.CO].
[15] F. Bianchini and A. Silvestri, Physical Review D **93**, 10.1103/physrevd.93.064026 (2016).
[16] W. J. Percival and M. White, MNRAS **393**, 297 (2009), arXiv:0808.0003.
[17] D. S. Swetz, P. A. R. Ade, M. Amiri, J. W. Appel, E. S. Battistelli, B. Burger, J. Chervenak, M. J. Devlin, S. R. Dicker, W. B. Doriese, et al., ApJS **194**, 41 (2011), arXiv:1007.0290 [astro-ph.IM].
[18] K. S. Dawson, D. J. Schlegel, C. P. Ahn, S. F. Anderson, É. Aubourg, S. Bailey, R. H. Barkhouser, J. E. Bautista, A. Beifiori, A. A. Berlind, et al., AJ **145**, 10 (2013), arXiv:1208.0022 [astro-ph.CO].
[19] N. Hand, G. E. Addison, E. Aubourg, N. Battaglia, E. S. Battistelli, D. Bizyaev, J. R. Bond, H. Brewington, J. Brinkmann, B. R. Brown, et al., Physical Review Letters **109**, 041101 (2012), arXiv:1203.4219 [astro-ph.CO].
[20] S. Padin, Z. Staniszewski, R. Keisler, M. Joy, A. A. Stark, P. A. R. Ade, K. A. Aird, B. A. Benson, L. E. Bleem, J. E. Carlstrom, et al., Appl. Opt. **47**, 4418 (2008).
[21] J. E. Carlstrom, P. A. R. Ade, K. A. Aird, B. A. Benson, L. E. Bleem, S. Busetti, C. L. Chang, E. Chauvin, H.-M. Cho, T. M. Crawford, et al., PASP **123**, 568 (2011), arXiv:0907.4445.
[22] B. Flaugher, H. T. Diehl, K. Honscheid, T. M. C. Abbott, O. Alvarez, R. Angstadt, J. T. Annis, M. Antonik, O. Ballester, L. Beaufore, et al., AJ **150**, 150 (2015), arXiv:1504.02900 [astro-ph.IM].
[23] B. Soergel, S. Flender, K. T. Story, L. Bleem, T. Giannantonio, G. Efstathiou, E. Rykoff, B. A. Benson, T. Crawford, S. Dodelson, et al., MNRAS **461**, 3172 (2016), arXiv:1603.03904.
[24] V. Calafut, P. Gallardo, E. Vavagiakis, S. Amodeo, S. Aiola, J. Austermann, N. Battaglia, E. Battistelli, J. Beall, R. Bean, et al., Physical Review D **104**, 10.1103/physrevd.104.043502 (2021).
[25] Z. Chen, P. Zhang, X. Yang, and Y. Zheng, Detection of pairwise ksz effect with desi galaxy clusters and planck (2021), arXiv:2109.04092 [astro-ph.CO].
[26] J. C. Hill, S. Ferraro, N. Battaglia, J. Liu, and D. N. Spergel, Physical Review Letters **117**, 10.1103/physrevlett.117.051301 (2016).
[27] E. Schaan, S. Ferraro, S. Amodeo, N. Battaglia, S. Aiola, J. E. Austermann, J. A. Beall, R. Bean, D. T. Becker, R. J. Bond, et al., Phys. Rev. D **103**, 063513 (2021), arXiv:2009.05557 [astro-ph.CO].
[28] N. Battaglia, Journal of Cosmology and Astroparticle Physics **2016** (08), 058–058.
[29] F. D. Bernardis, S. Aiola, E. Vavagiakis, N. Battaglia, M. Niemack, J. Beall, D. Becker, J. Bond, E. Calabrese, H. Cho, and et al., Journal of Cosmology and Astroparticle Physics **2017** (03), 008?008.
[30] E. Vavagiakis, P. Gallardo, V. Calafut, S. Amodeo, S. Aiola, J. Austermann, N. Battaglia, E. Battistelli, J. Beall, R. Bean, J. Bond, E. Calabrese, S. Choi, N. Cothard, M. Devlin, C. Duell, S. Duff, A. Duivenvoorden, J. Dunkley, R. Dunner, S. Ferraro, Y. Guan, J. Hill, G. Hilton, M. Hilton, R. Hložek, Z. Huber, J. Hubmayr, K. Huffenberger, J. Hughes, B. Koopman, A. Kosowsky, Y. Li, M. Lokken, M. Madhavacheril, J. McMahon, K. Moodley, S. Naess, F. Nati, L. Newburgh, M. Niemack, L. Page, B. Partridge, E. Schaan, A. Schillaci, C. Sifón, D. Spergel, S. Staggs, J. Ullom, L. Vale, A. V. Engelen, J. V. Lanen, E. Wollack, and Z. Xu, Physical Review D **104**, 10.1103/physrevd.104.043503 (2021).
[31] Planck Collaboration, N. Aghanim, Y. Akrami, M. Ashdown, J. Aumont, C. Baccigalupi, M. Ballardini, A. J. Banday, R. B. Barreiro, N. Bartolo, S. Basak, et al., A&A **641**, A6 (2020), arXiv:1807.06209 [astro-ph.CO].

[32] F. Bernardeau, S. Colombi, E. Gaztañaga, and R. Scoccimarro, Phys. Rep. 367, 1 (2002), astro-ph/0112551.
[33] A. Diaferio, R. A. Sunyaev, and A. Nusser, ApJL 533, L71 (2000), arXiv:astro-ph/9912117 [astro-ph].
[34] B. Soergel, A. Saro, T. Giannantonio, G. Efstathiou, and K. Dolag, Monthly Notices of the Royal Astronomical Society 478, 5320 (2018).
[35] R. Juszkiewicz, V. Springel, and R. Durrer, Astrophys. J. Lett. 518, L25 (1999), arXiv:astro-ph/9812387.
[36] R. K. Sheth, A. Diaferio, L. Hui, and R. Scoccimarro, Mon. Not. Roy. Astron. Soc. 326, 463 (2001), arXiv:astro-ph/0010137.
[37] S. Bhattacharya and A. Kosowsky, Phys. Rev. D 77, 083004 (2008), arXiv:0712.0034 [astro-ph].
[38] R. Keisler and F. Schmidt, ApJL 765, L32 (2013), arXiv:1211.0668 [astro-ph.CO].
[39] E.-M. Mueller, F. de Bernardis, R. Bean, and M. D. Niemack, Astrophys. J. 808, 47 (2015), arXiv:1408.6248 [astro-ph.CO].
[40] J. A. Sobrin, P. A. R. Ade, Z. Ahmed, A. J. Anderson, J. S. Avva, R. Basu Thakur, A. N. Bender, B. A. Benson, J. E. Carlstrom, F. W. Carter, et al., in Proc. SPIE, Proc. SPIE, Vol. 10708 (2018) p. 107081H, arXiv:1809.00032 [astro-ph.IM].
[41] D. Dutcher, L. Balkenhol, P. A. R. Ade, Z. Ahmed, E. Anderes, A. J. Anderson, M. Archipley, J. S. Avva, K. Aylor, P. S. Barry, et al., Phys. Rev. D 104, 022003 (2021), arXiv:2101.01684 [astro-ph.CO].
[42] M. R. Calabretta and E. W. Greisen, A&A 395, 1077 (2002), arXiv:astro-ph/0207413.
[43] K. K. Schaffer, T. M. Crawford, K. A. Aird, B. A. Benson, L. E. Bleem, J. E. Carlstrom, C. L. Chang, H. M. Cho, A. T. Crites, T. de Haan, et al., Astrophys. J. 743, 90 (2011), arXiv:1111.7245 [astro-ph.CO].
[44] E. S. Rykoff, E. Rozo, M. T. Busha, C. E. Cunha, A. Finoguenov, A. Evrard, J. Hao, B. P. Koester, A. Leauthaud, B. Nord, M. Pierre, R. Reddick, T. Sadibekova, E. S. Sheldon, and R. H. Wechsler, Astrophys. J. 785, 104 (2014), arXiv:1303.3562.
[45] E. S. Rykoff, B. P. Koester, E. Rozo, J. Annis, A. E. Evrard, S. M. Hansen, J. Hao, D. E. Johnston, T. A. McKay, and R. H. Wechsler, Astrophys. J. 746, 178 (2012), arXiv:1104.2089 [astro-ph.CO].
[46] A. Saro, S. Bocquet, E. Rozo, B. A. Benson, J. Mohr, E. S. Rykoff, M. Soares-Santos, L. Bleem, S. Dodelson, P. Melchior, et al., MNRAS 454, 2305 (2015), arXiv:1506.07814.
[47] E. Rozo, E. S. Rykoff, A. Abate, C. Bonnett, M. Crocce, C. Davis, B. Hoyle, B. Leistedt, H. V. Peiris, R. H. Wechsler, et al., MNRAS 461, 1431 (2016), arXiv:1507.05460 [astro-ph.IM].
[48] P. G. Ferreira, R. Juszkiewicz, H. A. Feldman, M. Davis, and A. H. Jaffe, The Astrophysical Journal 515, L1 (1999).
[49] A. Cavaliere and R. Fusco-Femiano, A&A 49, 137 (1976).
[50] J. Erler, M. E. Ramos-Ceja, K. Basu, and F. Bertoldi, Monthly Notices of the Royal Astronomical Society 484, 1988 (2019).
[51] J. Hartlap, P. Simon, and P. Schneider, A&A 464, 399 (2007), astro-ph/0608064.
[52] A. Klypin, G. Yepes, S. Gottlöber, F. Prada, and S. Heß, MNRAS 457, 4340 (2016), arXiv:1411.4001 [astro-ph.CO].
[53] A. J. Mead, T. Tröster, C. Heymans, L. Van Waerbeke, and I. G. McCarthy, Astronomy & Astrophysics 641, A130 (2020).
[54] I. G. McCarthy, J. Schaye, S. Bird, and A. M. C. Le Brun, Monthly Notices of the Royal Astronomical Society 465, 2936–2965 (2016).
[55] P. Behroozi, R. H. Wechsler, A. P. Hearin, and C. Conroy, Monthly Notices of the Royal Astronomical Society 488, 3143–3194 (2019).
[56] R. C. Kennicutt, Jr., ARA&A 36, 189 (1998), arXiv:astro-ph/9807187.
[57] T. McClintock, T. N. Varga, D. Gruen, E. Rozo, E. S. Rykoff, T. Shin, P. Melchior, J. DeRose, S. Seitz, et al., MNRAS 482, 1352 (2019), arXiv:1805.00039 [astro-ph.CO].
[58] S. Flender, L. Bleem, H. Finkel, S. Habib, K. Heitmann, and G. Holder, Astrophys. J. 823, 98 (2016), arXiv:1511.02843 [astro-ph.CO].
[59] L. E. Bleem, T. M. Crawford, B. Ansarinejad, B. A. Benson, S. Bocquet, J. E. Carlstrom, C. L. Chang, R. Chown, A. T. Crites, T. de Haan, et al., The Astrophysical Journal Supplement Series 258, 36 (2022).
[60] S. Grandis, J. J. Mohr, M. Costanzi, A. Saro, S. Bocquet, M. Klein, M. Aguena, S. Allam, J. Annis, B. Ansarinejad, et al., MNRAS 504, 1253 (2021), arXiv:2101.04984 [astro-ph.CO].
[61] J. Myles, D. Gruen, A. B. Mantz, S. W. Allen, R. G. Morris, E. Rykoff, M. Costanzi, C. To, J. DeRose, R. H. Wechsler, E. Rozo, T. Jeltema, E. R. Carrasco, A. Kremin, and R. Kron, MNRAS 505, 33 (2021), arXiv:2011.07070 [astro-ph.CO].
[62] H.-Y. Wu, M. Costanzi, C.-H. To, A. N. Salcedo, D. H. Weinberg, J. Annis, S. Bocquet, M. Elidaiana da Silva Pereira, J. DeRose, J. Esteves, et al., arXiv e-prints , arXiv:2203.05416 (2022), arXiv:2203.05416 [astro-ph.CO].
[63] D. E. Johnston, E. S. Sheldon, R. H. Wechsler, E. Rozo, B. P. Koester, J. A. Frieman, T. A. McKay, A. E. Evrard, M. R. Becker, and J. Annis, ArXiv e-prints (2007), arXiv:0709.1159.
[64] R. A. Sunyaev and Y. B. Zel'dovich, Comments on Astrophysics and Space Physics 2, 66 (1970).
[65] R. A. Sunyaev and Y. B. Zel'dovich, Comments on Astrophysics and Space Physics 4, 173 (1972).
[66] J. Delabrouille and J. F. Cardoso, Diffuse source separation in cmb observations (2007), arXiv:astro-ph/0702198 [astro-ph].
[67] R. Pordes, D. Petravick, B. Kramer, D. Olson, M. Livny, A. Roy, P. Avery, K. Blackburn, T. Wenaus, F. Würthwein, et al., in J. Phys. Conf. Ser., 78, Vol. 78 (2007) p. 012057.
[68] I. Sfiligoi, D. C. Bradley, B. Holzman, P. Mhashilkar, S. Padhi, and F. Wurthwein, in 2009 WRI World Congress on Computer Science and Information Engineering, Vol. 2 (2009) pp. 428–432.
[69] F. Pérez and B. E. Granger, Computing in Science & Engineering 9, 21 (2007), http://aip.scitation.org/doi/pdf/10.1109/MCSE.2007.53.
[70] J. D. Hunter, Computing In Science & Engineering 9, 90 (2007).
[71] P. Virtanen, R. Gommers, T. E. Oliphant, M. Haberland, T. Reddy, D. Cournapeau, E. Burovski, P. Peterson, W. Weckesser, J. Bright, et al., Nature Methods 17, 261 (2020).